\documentclass[journal]{IEEEtran}
\usepackage{amsfonts,amssymb}
\usepackage{color}
\usepackage{pifont}
\usepackage{hyperref}
\usepackage{afterpage}
\usepackage{multirow}
\usepackage{amsmath}
\usepackage{graphicx}
\usepackage{ntheorem}
 \usepackage{mathrsfs}
\newcommand{\fsquare}{\vrule height6pt width7pt depth1pt}   
\usepackage{cite}
\let\labelindent\relax
\usepackage{enumitem}
 \usepackage{cases}
 \usepackage{pbox}
\usepackage{subfigure}

  \usepackage{epsfig}

\newfont{\secfnta}{ptmb8t at 12pt}

\usepackage{makecell}
\usepackage{relsize}
\newcommand{\pfe}{\hfill\fsquare}
\newcommand{\1}[1]{{\bf 1}\left[#1\right]}

\newcommand{\bcap} {\hspace{2pt} \mathlarger{\cap}
\hspace{2pt}}

\newcommand{\bcup} {\hspace{2pt} \mathlarger{\cup}
\hspace{2pt}}

\newcommand*\mcapinn[2]{\vcenter{\hbox{$\mathsurround=0pt
\ifx\displaystyle#1\textstyle\else#1\fi\bigcap$}}}

\newcommand*\mcupinn[2]{\vcenter{\hbox{$
\bigcup$}}}

\newcommand{\bP}[1]{{\mathbb{P}}\left[{#1}\right]}
\newcommand{\bE}[1]{{\mathbb{E}}\left[{#1}\right]}

\newtheorem{lem}{Lemma}
\newtheorem{thm}{Theorem}

\newtheorem{proposition}{Proposition}

\begin{document}

\title{On Secure Communication in Sensor Networks under $q$-Composite Key Predistribution with Unreliable Links}



\author{Jun~Zhao,~\IEEEmembership{Member,~IEEE}
\thanks{Manuscript received February 15, 2017; revised July 2 and September 21, 2017; accepted November 23, 2017. Date of publication November 29, 2017; date of current version
November 29, 2017. This research was supported in part by the Cybersecurity Lab (CyLab) and Department of Electrical \& Computer Engineering
at Carnegie Mellon University, by Nanyang Technological University, and by Arizona State University. The associate editor coordinating the review of this paper
and approving it for publication was V. Aggarwal. (\emph{Corresponding author:
Jun Zhao}.)\newline \indent
The author was with Carnegie Mellon University, Pittsburgh, PA 15213, USA, and also with Arizona State University, Tempe, AZ 85281, USA. He is now with Nanyang Technological University, Singapore 639798. \newline \indent Color versions of one or more of the figures in this paper are available
online at http://ieeexplore.ieee.org.
\newline \indent Digital Object Identifier:
\newline \indent }}


\date{}

\maketitle


%




\begin{abstract}
 Many applications of wireless sensor networks (WSNs) require deploying sensors in hostile environments,
where an adversary may eavesdrop communications. To secure communications in WSNs, the $q$-composite key predistribution scheme has been proposed in the literature. In this paper, we
investigate secure $k$-connectivity in
WSNs operating under the $q$-composite scheme, in consideration of the unreliability of wireless
links. Secure $k$-connectivity ensures that any two sensors can find a path in between for secure communication, even when $k-1$ sensors fail. We present conditions on how to set the network parameters such that the network has secure $k$-connectivity asymptotically almost surely. The result is given in the form of a {sharp} zero--one law.


\end{abstract}


\section{Introduction}

\PARstart{W}{ireless} sensor networks (WSNs) enable a broad range
of applications including military surveillance, industrial monitoring, and home automation \cite{4625802}. When WSNs are   deployed in hostile environments,
  cryptographic mechanisms are needed to
secure  communications between sensors. Because the network topology is often unknown before deployment, the idea of key predistribution has been proposed to
protect sensor  communications~\cite{virgil}.

  Since Eschenauer and Gligor~\cite{virgil} introduced
the basic key predistribution scheme, key predistribution schemes
have been widely studied in the literature \cite{lu2008framework,yum2012exact,Zhu:2003:LES:948109.948120,kwon2009location,yaugan2016zero,bloznelis2013}. Among many key predistribution schemes,
the $q$-composite  scheme proposed by Chan
\emph{et al.} \cite{adrian} as an extension of the
Eschenauer--Gligor scheme~\cite{virgil} has received considerable interest  \cite{du2009routing,QcompTech14,JZISIT14,Rybarczyk,DiPietroTissec,DiPietroMeiManciniPanconesiRadhakrishnan2004,yagan_onoff} (the Eschenauer--Gligor scheme is the $q$-composite scheme in
the special case of $q=1$).
 The $q$-composite  scheme works as follows. For a WSN with $n$
sensors, prior to deployment, each sensor is independently assigned
$K_n$ different keys which are selected {uniformly at random}
from a pool $\mathcal {P}_n$ of $P_n$ distinct keys, where $K_n$ is referred to as the key ring size.  After deployment, any two sensors
establish a {secure} link in between {if
and only if} they share at least $q$ key(s) {and} the physical
link constraint between them is satisfied. $P_n$ and $K_n$ are
functions of $n$, with the natural condition $1 \leq q \leq
K_n \leq P_n$. Examples of physical link constraints include the
reliability of the transmission channel \cite{yagan_onoff,ZhaoYaganGligor,Yi06GLOBECOM,eletreby2017connectivity} and the requirement that the distance between two
sensors need to be close enough for direct communication~\cite{Gupta98criticalpower,yi2006asymptotic,4428764,Pishro}. The $q$-composite scheme
with $q\geq 2$ outperforms the   Eschenauer--Gligor scheme with
$q=1$ in terms of the strength against small-scale sensor capture while trading off increased vulnerability in the face of
large-scale attacks \cite{adrian}.


 In this
paper, we investigate secure $k$-connectivity in
WSNs employing the $q$-composite key predistribution scheme with
the
physical link constraint represented by the \emph{on}/\emph{off} channel model comprising independent channels which are
either \emph{on} or \emph{off}. Secure $k$-connectivity ensures that any two sensors can find a path in between for secure communication, even when any $k-1$ sensors fail and are deleted from the network topology. The on/off channel model captures {the unreliability of wireless links} due
to physical barriers between sensors or  harsh environmental conditions   impairing communications \cite{wang2011cooperation,yavuz2017k,zhao2011fundamental}.
Our results are given in the form of a \emph{sharp} zero--one law, meaning that the
network is securely $k$-connected asymptotically almost surely (\textit{a.a.s.}) under certain parameter conditions
and does not have secure $k$-connectivity  \textit{a.a.s.} if parameters are slightly changed,
where an event happens \textit{a.a.s.} if its probability converges to $1$ over a sequence of sets (i.e., in this paper, as the number of sensors tends to infinity).
In the asymptotic sense, the \mbox{zero--one} law specifies the critical scaling of the model parameters in terms of secure $k$-connectivity. Despite being asymptotic, such a critical scaling provides useful insights to understand secure WSNs.
%
In a secure WSN, to increase the probability of $k$-connectivity,
it is often required to enlarge the number of keys in each
sensor's memory. However, since sensors are expected to have
 limited memory, it is desirable for key
distribution schemes to have low memory requirements
\cite{virgil,yagan,adrian,zhao2017ITq}. Therefore, it is important to establish a zero--one law in order to carefully
dimension the $q$-composite key predistribution scheme for secure communications between sensors.

%
%

We organize the rest of the paper as follows. After Section~\ref{sec:SystemModel} describes the system model,
 Section~\ref{sec:res} presents the results. We survey related work in Section~\ref{related}.
 Sections~\ref{sec:ProofTheoremNodeIsolation} and~\ref{sec:lem_Gq_no_isolated_but_not_conn} are devoted to proving the results. 
 Finally,  we conclude the paper in Section \ref{sec:Conclusion}. 

 \section{System Model}
\label{sec:SystemModel}

The studied WSN consists of $n$ sensors,
employs the $q$-composite key predistribution scheme, and works under
the {on/off} channel model.
We will explain that the graph representing the studied WSN  is an intersection
of two distinct types of random graphs. The intertwining
of random graphs makes the analysis challenging.

 We use a node set $\mathcal{V}_n = \{v_1,
v_2, \ldots, v_n \}$ to represent the $n$ sensors (the terms sensor and node are interchangeable
in this paper). For each node
$v_i \in \mathcal{V}_n$, let the set of its $K_n$ different keys be $S_i$. According to the $q$-composite key predistribution scheme,  $S_i$ is uniformly distributed among all
$K_n$-size subsets of a key pool $\mathcal {P}_n$ of $P_n$ keys.


The $q$-composite key predistribution scheme is
 modeled by a uniform $q$-intersection graph~\cite{bloznelis2013}
denoted by $G_q(n, K_n,P_n)$. In such a graph defined on the node set
$\mathcal{V}_n$, any two distinct nodes $v_i$ and $v_j$ have
an edge in between if and only if they
share at least $q$ key(s) (an event denoted by $\Gamma_{ij}$) . With $|A|$ being the
cardinality of a set $A$, event $\Gamma_{ij}$ is given by  $\big[ |S_i
\bcap S_j | \geq q \big]$.

Under the {on/off} channel model, each node-to-node channel is
independently {\em on} with probability ${p_n} $ and {\em off} with
probability $(1-{p_n})$, where ${p_n}$ is a function of $n$ with
$0<{p_n}\leq 1$. Letting ${L}_{i j}$ be the event that the channel
between distinct nodes $v_i$ and $v_j$ is {\em on}, we have
$\bP{L_{ij}} = {p_n}$, where {$\mathbb{P}[\mathcal {E}]$ denotes the
probability that an event $\mathcal {E}$ happens, throughout the paper.}
The network topology under the {on/off} channel model is given by an Erd\H{o}s--R\'enyi
graph $G(n, {p_n})$~\cite{citeulike:4012374} with the node set being
$\mathcal{V}_n$ and the edge set specified by $L_{ij}$.

Finally, we use $\mathbb{G}_q(n, K_n, P_n,
{p_n})$ to model the $n$-node WSN operating under the
$q$-composite scheme and the on/off channel
model. In graph
$\mathbb{G}_q(n, K_n, P_n,
{p_n})$ defined on the node set
$\mathcal{V}_n$, there exists an edge between nodes $v_i$ and
$v_j$ (an event denoted by $E_{ij}$) if and only if events $\Gamma_{ij}$ and $L_{ij}$ both happen. We have $E_{ij}  = \Gamma_{ij} \cap L_{ij}$.
Clearly, the edge set of $\mathbb{G}_q(n, K_n, P_n,
{p_n})$ is the intersection of the edge sets of $G_q(n, K_n, P_n) $ and $G(n, {p_n})$, and these graphs are all defined on the vertex set $\mathcal{V}_n$. Then
   $\mathbb{G}_q(n, K_n, P_n,
{p_n})$ can be seen as the intersection
of $G_q(n, K_n, P_n)$ and $G(n, {p_n})$; i.e.,
\begin{equation}
\mathbb{G}_q(n, K_n, P_n,
{p_n}) = G_q(n, K_n, P_n) \cap G(n, {p_n}).
\nonumber
\end{equation}

In Erd\H{o}s--R\'enyi
graph $G (n, p_n)$, all edges are independent of each other. However, in graph $G_q(n, K_n, P_n)$, the edges are not independent since the events that different pairs of three nodes share $q$ key(s) are not independent. A recent work \cite{bloznelis2013} demonstrates different behavior of $G_q(n, K_n, P_n)$ and $G (n, p_n)$ in terms of clustering
coefficient.

Throughout the
paper, $q$ and $k$ are arbitrary positive integers and do not scale with
$n$.
We define $s(K_n, P_n,q)$ as the probability that two different nodes
share at least $q$ key(s) and $t (K_n, P_n,q, {p_n})$ as the probability that two
distinct nodes have a secure link in $\mathbb{G}_q(n,
K_n, P_n, {p_n})$. We often write $s(K_n, P_n,q)$ and $t (K_n, P_n,q, {p_n})$
as $s_n$ and $t_n$ respectively for simplicity.
 Clearly,
$s_n $ and $t_n$ are the edge probabilities in graphs
$G_q(n, K_n, P_n)$ and $\mathbb{G}_q(n,
K_n, P_n, {p_n})$, respectively.  From $E_{ij} = L_{ij} \cap \Gamma_{ij}$ and the
independence of $L_{ij} $ and $ \Gamma_{i j} $, we obtain
\begin{align}
{t_n}  & =  \mathbb{P} [E_{i j} ]  =  \mathbb{P} [L_{ij} ]
\cdot \mathbb{P} [\Gamma_{i j} ] =  {p_n}\cdot
s_n. \label{eq_pre}
\end{align}

 By definition, $s_n$ is determined through
\begin{align}
s_n & =  \mathbb{P} [\Gamma_{i j} ] = \sum_{u=q}^{K_n}
  \mathbb{P}[|S_{i} \cap S_{j}| = u] ,  \label{psq1}
\end{align}
where it holds for $P_n \geq 2K_n$ that
\begin{align}
\mathbb{P}[|S_{i} \cap S_{j}| = u] & =
\frac{\binom{K_n}{u}\binom{P_n-K_n}{K_n-u}}{\binom{P_n}{K_n}},
\quad\text{for } u =1,2,\ldots, K_n,
\label{u3}
\end{align}
which along with (\ref{eq_pre}) and (\ref{psq1}) induce that under $P_n \geq 2K_n$,
\begin{align}
t_n = {p_n}\cdot  \sum_{u=q}^{K_n}
\frac{\binom{K_n}{u}\binom{P_n-K_n}{K_n-u}}{\binom{P_n}{K_n}}. \label{u4}
\end{align}

\section{The Results} \label{sec:res}

We now present the results.
The
natural logarithm function is given by $\ln$.   
 We use the standard
asymptotic notation $o(\cdot), \omega(\cdot), O(\cdot), \Omega(\cdot),
\Theta(\cdot), \sim$ in~\cite[Footnote 1]{zhao2017ITq}. These symbols and all other limits are understood with $n \to  \infty$.

Theorem \ref{thm:OneLaw+NodeIsolation} below presents a sharp zero--one law for $k$-connectivity in a  graph $\mathbb{G}_q(n, K_n,P_n, {p_n})$. In the secure sensor network modeled by $\mathbb{G}_q(n, K_n,P_n, {p_n})$, $k$-connectivity enables any two sensors to have secure communication either directly or through the help of relaying nodes, even when any $k-1$ sensors are removed from the network.

\begin{thm}  \label{thm:OneLaw+NodeIsolation}

For a graph $\mathbb{G}_q(n, K_n,P_n, {p_n})$, with a sequence $\alpha_n$ defined through
\begin{align}
t_n  & = \frac{\ln  n + (k-1)\ln \ln n  +
 {\alpha_n}}{n},   \label{eq:scalinglaw}
\end{align}
where $t_n$ denoting the edge probability of $\mathbb{G}_q(n, K_n,P_n, {p_n})$ is given by (\ref{u4}), then it holds under $ P_n =
\Omega(n)$ and $\frac{{K_n}^2}{P_n} = o(1)$ that
\begin{align}
\lim_{n \rightarrow \infty }  \mathbb{P} \bigg[
\begin{array}{c}
\mathbb{G}_q(n, K_n,P_n, {p_n}) \\
\mbox{is $k$-connected.}
\end{array}
\bigg] \nonumber
\end{align}
\begin{subnumcases}{=}  0,\quad \text{if  }\lim_{n \to \infty}{\alpha_n}   =   - \infty, \label{thm-con-eq-0} \\
1,\quad \text{if  }\lim_{n \to \infty}{\alpha_n}    =  \infty. \label{thm-con-eq-1}
\end{subnumcases}
\end{thm}

Theorem \ref{thm:OneLaw+NodeIsolation} presents a strong zero--one law for $k$-connectivity in graph $\mathbb{G}_q(n, K_n,P_n, {p_n})$, where a critical scaling of $t_n$ can be set as $\frac{\ln  n + (k-1)\ln \ln n +c}{n}$ with any constant $c$. In addition, the conditions $P_n =
\Omega(n)$ and $\frac{{K_n}^2}{P_n}=o(1)$ in Theorem \ref{thm:OneLaw+NodeIsolation} are reasonable, since it is expected
\cite{adrian,DiPietroMeiManciniPanconesiRadhakrishnan2004,virgil,yum2012exact}
that for security purposes, the key pool size $P_n$ is at least on the order of the node number $n$, and is much larger than the number $K_n$ of keys on each sensor. For example, for $n$  between 1000 and 10000, Di~Pietro~\textit{et~al.}~\cite{DiPietroTissec} find that a suitable choice is to set $P_n$ as $\frac{n \ln n}{32}$ and set $K_n$ as $\ln n$.



 \section{Related Work} \label{related}

We now compare Theorem \ref{thm:OneLaw+NodeIsolation} in this paper with related results \cite{QcompTech14,ZhaoYaganGligor,zhao2017ITq,yagan_onoff} in the literature. After the detailed comparison, we discuss more related work.

\textbf{Comparison with \cite{QcompTech14}.} Recently, \cite[Theorem 1]{QcompTech14} presents the result on the probability of minimum degree being at least $k$ in $\mathbb{G}_q(n,
K_n, P_n, {p_n})$. An extension to $k$-connectivity is also given in~\cite{QcompTech14}. Below, we first explain that the results for $k$-connectivity in this paper are stronger than the $k$-connectivity results in \cite{QcompTech14}, and then show that the proof techniques in this paper are  more advanced than those  in \cite{QcompTech14}.

 To ensure $k$-connectivity (i.e., the \mbox{one-law}  part), we need $t_n  = \frac{\ln  n  + (k-1)\ln \ln n   +
 {\alpha_n}}{n}$ with
$\lim_{n \to \infty}\alpha_n = \infty$. Then the requirement on the key ring size $K_n$ (i.e., the number of keys on each sensor) in this paper for $k$-connectivity is $K_n = \Omega\Big( n^{\frac{1}{2}-\frac{1}{2q}}(\ln n )^{\frac{1}{2q}} {p_n} ^{-\frac{1}{2q}} \Big)$ according to Lemma \ref{lemboundKn}-Property (ii) below, while the requirement on the key ring size $K_n$ in \cite{QcompTech14} for $k$-connectivity satisfies $K_n =  \omega\left(n^{1-\frac{1}{q}}(\ln n )^{1+\frac{1}{q}} {p_n} ^{-\frac{1}{q}} \right)$ according to Lemma \ref{lemboundKn}-Property (iii) below, although the $k$-connectivity result in \cite{QcompTech14} mentions $K_n = \Omega(n^{\epsilon})$  for a positive constant
 $\epsilon$ (note that $  \omega\left(n^{1-\frac{1}{q}}(\ln n )^{1+\frac{1}{q}} {p_n} ^{-\frac{1}{q}} \right)$ for $q\geq 2$ satisfies $\Omega(n^{\epsilon})$ for $\epsilon \leq 1-\frac{1}{q}$). Then we see that the order $n^{1-\frac{1}{q}}(\ln n )^{1+\frac{1}{q}} {p_n} ^{-\frac{1}{q}}$ of the minimal $K_n$ in \cite{QcompTech14} is more than the square of the order $n^{\frac{1}{2}-\frac{1}{2q}}(\ln n )^{\frac{1}{2q}} {p_n} ^{-\frac{1}{2q}}$ of the minimal $K_n$ in this paper, given ${n^{1-\frac{1}{q}}(\ln n )^{1+\frac{1}{q}} {p_n} ^{-\frac{1}{q}}}\Big/{\left(n^{\frac{1}{2}-\frac{1}{2q}}(\ln n )^{\frac{1}{2q}} {p_n} ^{-\frac{1}{2q}}\right)^2} = \ln n $.

Lemma \ref{lemboundKn} below presents the requirement on the key ring size $K_n$ in this paper and \cite{QcompTech14} for $k$-connectivity.

\begin{lem} \label{lemboundKn}
Under $ P_n = \omega(1)$, if the sequence $\alpha_n $ defined by (\ref{eq:scalinglaw}) satisfies either
$\lim_{n \to \infty}\alpha_n = \infty$ or $|\alpha_n| = o(\ln n) $, then
\begin{itemize}
\item[(i)] we have $K_n = \Omega\Big(n^{-\frac{1}{2q}}(\ln n )^{\frac{1}{2q}} {p_n} ^{-\frac{1}{2q}}  \cdot \sqrt{P_n} \hspace{2pt} \Big)$;
\item[(ii)]  if $ P_n =
\Omega(n)$ (a condition of Theorem \ref{thm:OneLaw+NodeIsolation} in this paper), we have $K_n = \Omega\Big( n^{\frac{1}{2}-\frac{1}{2q}}(\ln n )^{\frac{1}{2q}} {p_n} ^{-\frac{1}{2q}} \Big)$;
\item[(iii)]  if $\frac{K_n}{P_n} = o\left(\frac{1}{n\ln n}\right)$ (a condition in the discussion  of \cite{QcompTech14}), we have $K_n =  \omega\left(n^{1-\frac{1}{q}}(\ln n )^{1+\frac{1}{q}} {p_n} ^{-\frac{1}{q}} \right)$.\end{itemize}
\end{lem}

\noindent \textbf{Proof of Lemma \ref{lemboundKn}:}

\textit{Proving property (i):}

 Given (\ref{eq:scalinglaw}) (i.e., $t_n  = \frac{\ln  n  + (k-1)\ln \ln n   +
 {\alpha_n}}{n}$), we know from either
$\lim_{n \to \infty}\alpha_n = \infty$ or $|\alpha_n| = o(\ln n) $ that $t_n    = \Omega\big(\frac{\ln  n}{n}\big)$. Note that when $|\alpha_n| = o(\ln n) $, we   have the stronger result $t_n    = \Theta\big(\frac{\ln  n}{n }\big)$, but we can still write $t_n    = \Omega\big(\frac{\ln  n}{n }\big)$. Then $t_n    = \Omega\big(\frac{\ln  n}{n }\big)$ and $t_n  = p_n s_n$ of (\ref{eq_pre}) imply
\begin{align}
s_n = \Omega\bigg(\frac{\ln  n}{n p_n}\bigg).  \label{pn1tnsnlnng}
\end{align}
Note that the $q$-composite scheme enforces the natural condition $1 \leq q \leq
K_n \leq P_n$. Recently, in {\cite[Lemma 6]{bloznelis2013},}
 Bloznelis shows  $s_n \leq \frac{\big[\binom{K_n}{q}\big]^2}{\binom{P_n}{q}}$, which further means
 \begin{align}
s_n \leq \frac{({{K_n}^q}/{q!} )^2}{{{(P_n-q)}^q}/{q!} } = \frac{1}{q!}\bigg(\frac{{K_n}^2}{P_n-q}\bigg)^q\sim \frac{1}{q!}\bigg(\frac{{K_n}^2}{P_n}\bigg)^q, \label{pn1tnsnlnng2}
\end{align}
 where the last step uses $P_n = \omega(1)$.

We use (\ref{pn1tnsnlnng}) and (\ref{pn1tnsnlnng2}) to derive $\frac{{K_n}^2}{P_n} = \Omega\Big(\big(\frac{\ln  n}{n p_n}\big)^{\frac{1}{q}} \Big)$, which  implies
\begin{align}
 & K_n = \sqrt{\textstyle{\Omega\Big(\big(\frac{\ln  n}{n p_n}\big)^{\frac{1}{q}} \Big) \cdot P_n}} = \Omega\Big(n^{-\frac{1}{2q}}(\ln n )^{\frac{1}{2q}} {p_n} ^{-\frac{1}{2q}}  \cdot \sqrt{P_n} \hspace{2pt} \Big) . \nonumber
\end{align}



\textit{Proving property (ii):}

\noindent
We use the condition $P_n = \Omega(n)$ of property (ii) and the result \mbox{$K_n = \Omega\Big(n^{-\frac{1}{2q}}(\ln n )^{\frac{1}{2q}} {p_n} ^{-\frac{1}{2q}}  \cdot \sqrt{P_n} \hspace{2pt} \Big)$} of property (i) to obtain $K_n = \Omega\Big(n^{-\frac{1}{2q}}(\ln n )^{\frac{1}{2q}} {p_n} ^{-\frac{1}{2q}}  \cdot \sqrt{n} \hspace{2pt} \Big) = \Omega\Big( n^{\frac{1}{2}-\frac{1}{2q}}(\ln n )^{\frac{1}{2q}} {p_n} ^{-\frac{1}{2q}} \Big). $

\textit{Proving property (iii):}

\noindent
We use the condition $\frac{K_n}{P_n} = o\left(\frac{1}{n\ln n}\right)$ of property (iii) and the result \mbox{$K_n = \Omega\Big(n^{-\frac{1}{2q}}(\ln n )^{\frac{1}{2q}} {p_n} ^{-\frac{1}{2q}}  \cdot \sqrt{P_n} \hspace{2pt} \Big)$} of property (i) to derive $K_n   = \Omega\Big( n^{-\frac{1}{2q}}(\ln n )^{\frac{1}{2q}} {p_n} ^{-\frac{1}{2q}}  \cdot \sqrt{\omega(K_n n\ln n)}  \Big)  = \sqrt{K_n}   \cdot   \omega \left( n^{\frac{1}{2}-\frac{1}{2q}}(\ln n )^{\frac{1}{2}+\frac{1}{2q}} {p_n} ^{-\frac{1}{2q}} \right)$, which further implies $K_n =  \omega\left(n^{1-\frac{1}{q}}(\ln n )^{1+\frac{1}{q}} {p_n} ^{-\frac{1}{q}} \right)$.
\pfe



We have explained above that the  $k$-connectivity results   in this paper are stronger than the $k$-connectivity results in \cite{QcompTech14}. We now discuss the underlying reason: the proof techniques in this paper are  better than those in \cite{QcompTech14}.  Specifically, the challenges for $k$-connectivity analysis in graph $\mathbb{G}_q(n, K_n,P_n, {p_n}) $ result from the dependencies between the edges as well as the intertwining
between different random graphs $G_q(n, K_n, P_n)$ and $G(n, {p_n})$ in the graph intersection $\mathbb{G}_q(n, K_n,P_n, {p_n}) = G_q(n, K_n, P_n) \cap G(n, {p_n})$. The edge dependencies in $\mathbb{G}_q(n, K_n,P_n, {p_n}) $ exist
since the events that different pairs of three nodes share $q$ key(s) are not independent. To address the above challenges for $k$-connectivity analysis, we carefully analyze the graph structure of $\mathbb{G}_q(n, K_n,P_n, {p_n}) $ and present a direct proof. In contrast, \cite{QcompTech14} provides an indirect proof by building the relationship between $\mathbb{G}_q(n, K_n,P_n, {p_n}) $ and another simpler random graph where the above dependencies between the edges are canceled out. As already discussed above, the  $k$-connectivity results derived from our direct proof are much stronger than those derived from the indirect proof in \cite{QcompTech14}.


\textbf{Comparison with \cite{yagan_onoff,ZhaoYaganGligor}.}
As detailed in Section
\ref{sec:SystemModel}, the graph model $\mathbb{G}_q(n, K_n,P_n, {p_n}) = G_q(n, K_n, P_n) \cap G(n,p_n)$ studied in this paper represents the topology of a secure sensor network employing the $q$-composite key predistribution scheme
\cite{virgil} under the on/off channel model.
 When $q=1$, graph $\mathbb{G}_q(n, K_n,P_n, {p_n}) $ reduces to $\mathbb{G}_1(n, K_n,P_n, {p_n}) $, which models the topology of a secure sensor network employing the Eschenauer--Gligor key predistribution scheme under the on/off channel model. For graph $\mathbb{G}_1(n, K_n,P_n, {p_n}) $, Ya\u{g}an \cite{yagan_onoff} presents a zero--one law for
connectivity, while Zhao~\textit{et~al.}~\cite{ZhaoYaganGligor} extend the result to $k$-connectivity.
 Below we compare \cite{yagan_onoff,ZhaoYaganGligor} and this paper. {First}, our result is for general $q$, while the results of \cite{yagan_onoff,ZhaoYaganGligor} are only for the case of $q$ being $1$.
{Second}, our result eliminates Ya\u{g}an's condition on
the existence of
$\lim_{n\to \infty}({p_n\iffalse_{on}\fi}\ln n)$, and eliminates \cite{ZhaoYaganGligor}'s condition  that either
there exists $\epsilon>0$ such that $ s(K_n, P_n, 1) p_n   n > \epsilon$
holds for all $n$ sufficiently large {or} $\lim_{n \to \infty}  [s(K_n, P_n, 1) p_n n]  =0$.

\textbf{Comparison with \cite{zhao2017ITq}.} Recently, \cite{zhao2017ITq} studies connectivity of secure sensor networks under the $q$-composite key predistribution scheme, when  two sensors sharing $q$ key(s)  also need to satisfy constraints of the well-known disk model \cite{Gupta98criticalpower,DiPietroMeiManciniPanconesiRadhakrishnan2004,4428764,Pishro} for  direct communication; i.e., two sensors have to be within certain distance to establish a link. In addition to the disk model, \cite{zhao2017ITq} also considers the combination of the disk model and the on/off channel model. Although the networks in \cite{zhao2017ITq} represent more complex graphs, the results of \cite{zhao2017ITq} are just for connectivity (not for $k$-connectivity), and just about
 one-laws (not about zero--one laws). In fact, even if zero-laws are added, \cite{zhao2017ITq} presents weaker granularity of zero--one laws compared with this paper, as explained below.
 We now present the zero--one law under the disk model in detail. In secure sensor networks
employing the $q$-composite scheme under the disk model where $n$ sensors
 are independently and uniformly deployed in a network field $\mathcal{A}$ of unit area, two sensors
have a secure link in between if and only if (i) they share at least
$q$ keys, {and} (ii) they have a distance no greater than
$r_n$. The former constraint results in a uniform $q$-intersection graph $G_q(n, K_n, P_n)$ discussed before, whereas the latter constraint induces a random geometric
 graph $G_{RGG}(n, r_n,\mathcal{A})$, so the network is modeled by the intersection $ G_q(n, K_n, P_n) \cap G_{RGG}(n, r_n,\mathcal{A})$.  If   the network field $\mathcal{A}$ is a unit torus so that the boundary effect~\cite{mao2013connectivity,JZNodeCapture} is ignored, the one-law in \cite{zhao2017ITq} and its zero-law extension \cite{JZNodeCapture} present the following results: under $K_n = \omega(\ln n)$, $K_n  = o\Big(\min\big\{\sqrt{P_n}, \frac{P_n}{n}\big\}\Big)$,
$r_n = o(1)$ and
\begin{equation}
s(K_n, P_n, q) \cdot \pi {r_n}^2 \sim \frac{c\ln n}{n}
\label{eq:scaling_jz_comp_osyafr}
\end{equation}
for a positive constant $c$,
  graph
$ G_q(n, K_n, P_n) \cap G_{RGG}(n, r_n,\mathcal{A})$ is disconnected \textit{a.a.s.} if $c<1$ and connected \textit{a.a.s.} if $c>1$. Note that although the results in \cite{zhao2017ITq} actually use $  \frac{1}{q!}\big(\frac{{K_n}^2}{P_n}\big)^q  \cdot \pi {r_n}^2 $ in (\ref{eq:scaling_jz_comp_osyafr}), we replace it by $s(K_n, P_n, q) \cdot \pi {r_n}^2$ for better comparison given $s_n
 \sim \frac{1}{q!}\big(\frac{{K_n}^2}{P_n}\big)^q$. If the boundary effect of network fields is considered; for example, if the network field $\mathcal{A}$ is a unit square with the boundary effect, then the results  need to replace $\frac{c\ln n}{n}  $ in (\ref{eq:scaling_jz_comp_osyafr}) by  $c \times \max \big\{ \frac{\ln n +   \ln [1/s(K_n, P_n, q) ]}{n}   ,~ \frac{4  \ln [1/s(K_n, P_n, q) ]}{n}  \big\} $. Hence, the results considering the boundary effect under the disk model are complex and   different from those under the on/off channel model. Below  we discuss  only the case of ignoring the boundary effect of network fields, in order to compare the disk model with the on/off channel model.

 From Theorem \ref{thm:OneLaw+NodeIsolation}, under $P_n = \Omega(n)$ and $\frac{{K_n}^2}{P_n} = o(1)$, with $\alpha_n$ defined through
\begin{equation}
 s(K_n, P_n, q) \cdot p_n  = \frac{\ln  n + (k-1)\ln \ln n   +
 {\alpha_n}}{n},  \label{eq:scaling_jz_comp_osyaf2rst}
 \end{equation}
  graph $\mathbb{G}_q(n, K_n,P_n, {p_n})$ (i.e., $ G_q(n, K_n, P_n) \cap G(n, {p_n})$) is not $k$-connected \textit{a.a.s.} if $\lim_{n \to \infty}{\alpha_n}
 = -\infty$ and $k$-connected \textit{a.a.s.} if $\lim_{n \to \infty}{\alpha_n}
 = \infty$.

  As discussed above, the connectivity results under the disk model ignoring the boundary effect use the scaling $\frac{c\ln  n}{n}$ for $c<1$ or $c>1$, whereas the   scaling  in this paper is $\frac{\ln  n + (k-1) \ln \ln  n + \alpha_n}{n}$ for $\lim_{n \to \infty}{\alpha_n}   = - \infty$ or $\lim_{n \to \infty}{\alpha_n}   =  \infty$ (for $k=1$, the   scaling  in this paper becomes $\frac{\ln  n  + \alpha_n}{n}$). The   scaling $\frac{\ln  n + (k-1) \ln \ln  n + \alpha_n}{n}$  in this paper ($\frac{\ln  n  + \alpha_n}{n}$ for $k=1$) is more fine-grained than the   scaling $\frac{c\ln  n}{n}$ in \cite{zhao2017ITq} because
 a deviation of $\alpha_n= \pm \,\Omega(\ln n)$
is required to get the zero--one law in the form of $\frac{c\ln  n}{n}$ for $c<1$ or $c>1$, whereas in $\frac{\ln  n  + \alpha_n}{n}$,
it suffices to have an unbounded deviation, e.g., even $\alpha_n = \pm \ln\ln \cdots \ln n$ will do.
Put differently, when $k=1$, the   scaling  $\frac{\ln  n  + \alpha_n}{n}$ in this paper covers the case of $c=1$ in $\frac{c\ln  n}{n}$, and shows that in this case, the graph could be connected or disconnected  \textit{a.a.s.}, depending on the limit
of ${\alpha_n}$. Although this paper and \cite{zhao2017ITq} use different scalings, we note that graph
$ G_q(n, K_n, P_n) \cap G(n, {p_n}) $ and $ G_q(n, K_n, P_n) \cap G_{RGG}(n, r_n,\mathcal{A})$
have   similar connectivity properties when they are {\em matched}
through edge probabilities so that $  s(K_n, P_n, q) \cdot p_n $ in (\ref{eq:scaling_jz_comp_osyaf2rst}) is equivalent with
$s(K_n, P_n, q) \cdot \pi {r_n}^2$ in (\ref{eq:scaling_jz_comp_osyafr}) (i.e. when $p_n $ and $\pi {r_n}^2$ are the same).

We now explain that the results for $k$-connectivity under the on/off channel model in this paper are stronger than those under the disk model in \cite{zhao2017ITq}. Specifically, this paper  considers $K_n = \Omega\Big( n^{\frac{1}{2}-\frac{1}{2q}}(\ln n )^{\frac{1}{2q}} {p_n} ^{-\frac{1}{2q}} \Big)$ from Lemma \ref{lemboundKn}-Property (ii), while \cite{zhao2017ITq} requires $K_n =  \omega\left(n^{1-\frac{1}{q}}(\ln n )^{\frac{1}{q}} (\pi {r_n}^2 ) ^{-\frac{1}{q}} \right)$ according to Footnote \ref{footnoteKn} below\footnote{Although the results in \cite{zhao2017ITq} mention $K_n = \omega(\ln n)$, we show that the required condition satisfies $K_n =  \omega\Big(n^{1-\frac{1}{q}}(\ln n )^{\frac{1}{q}} (\pi {r_n}^2 ) ^{-\frac{1}{q}} \Big)$. From (\ref{eq:scaling_jz_comp_osyafr}), to ensure connectivity, \cite{zhao2017ITq} needs $s(K_n, P_n, q) \cdot \pi {r_n}^2 \sim \frac{c\ln n}{n} $ with $c>1$, which with the condition $r_n = o(1)$ implies $s(K_n, P_n, q)  = \Omega\big(\frac{\ln  n}{n \cdot \pi {r_n}^2 }\big)$. Then similar to the proof of Lemma \ref{lemboundKn}-Property (i) (we just replace $p_n$ therein by $\pi {r_n}^2$), we derive $K_n = \Omega\Big(n^{-\frac{1}{2q}}(\ln n )^{\frac{1}{2q}} {p_n} ^{-\frac{1}{2q}}  \cdot \sqrt{P_n} \hspace{2pt} \Big)$, which along with $K_n  = o\big(\frac{P_n}{n}\big)$ (a condition in \cite{zhao2017ITq}) implies $K_n =  \omega\Big(n^{1-\frac{1}{q}}(\ln n )^{\frac{1}{q}} (\pi {r_n}^2 ) ^{-\frac{1}{q}} \Big)$. \label{footnoteKn}}. In other words, when $p_n$ and $\pi {r_n}^2$ are the same, the order for minimal $K_n$ in \cite{zhao2017ITq} is roughly the square of the order for minimal $K_n$ in this paper.

 In addition to the above differences, similar to \cite{QcompTech14}, the reference \cite{zhao2017ITq} also uses an indirect proof by building the relationship between the studied graph and another simpler random graph where the dependencies between the edges are canceled out. In contrast, this paper's proof is based on an direct analysis of the graph structure.

\textbf{Connectivity of graph ${G}_q(n, K_n, P_n)$.}
Graph ${G}_q(n, K_n, P_n)$ models
the topology of a secure sensor network with the $q$-composite key predistribution   under full visibility, which
means that any node pair have active channels in between so the only requirement for a secure link
is the sharing of at least $q$ keys.
For $G_q(n, K_n, P_n)$, Bloznelis and {\L}uczak
\cite{Perfectmatchings}
  have   derived a zero--one law for connectivity, while an extension to $k$-connectivity has been given by Bloznelis and Rybarczyk
\cite{bloznelis2013,Bloznelis201494}. Other properties of $G_q(n, K_n, P_n)$ are also considered in the literature~\cite{Rybarczyk}.  When $q=1$,   ${G}_1(n, K_n, P_n)$ models
the topology of a secure sensor network with the Eschenauer--Gligor key predistribution scheme under full visibility. For ${G}_1(n, K_n, P_n)$, its connectivity
  has been investigated extensively \cite{ryb3,r1,yagan,DiPietroTissec,DiPietroMeiManciniPanconesiRadhakrishnan2004}.

\textbf{Connectivity of Erd\H{o}s--R\'{e}nyi graph $G(n,p_n)$.}
  Erd\H{o}s and R\'{e}nyi \cite{citeulike:4012374} 
 introduce the random graph model $G(n,p_n)$ defined on
a node set with size $n$ such that an edge between any two nodes
exists with probability $p_n$ {independently} of all other
edges. Graph $G(n,p_n)$ models the topology induced by a sensor
network under the on/off channel model (when geometric constraints for transmissions are not considered). From \cite{citeulike:4012374}'s result and our Theorem \ref{thm:OneLaw+NodeIsolation}, Erd\H{o}s--R\'enyi graph $G(n,p_n')$ and graph
$\mathbb{G}_q(n, K_n,P_n, {p_n}) $
have   similar connectivity properties when they are {\em matched}
through edge probabilities (i.e. when $ p_n'$ equals $t_n$ in the left hand side of (\ref{eq:scalinglaw})).

\textbf{Connectivity of wireless networks under the disk model or its variants.}
Many connectivity studies~\cite{Gupta98criticalpower,yi2006asymptotic,DiPietroMeiManciniPanconesiRadhakrishnan2004,4428764,Pishro} of wireless networks use the disk model, where two nodes have to be within certain distance for direct communication. For the node distribution, two common models are as follows: 1) the uniform node distribution, where nodes
 are uniformly and independently deployed in a network field, and 2) the Poisson node distribution, where   nodes are distributed according to a Poisson point process. Results under these two   distributions are often shown to be equivalent since they can be connected via Chebyshev's inequality, which bounds the number of nodes in a Poisson point process; see    (de)Poissonization   in \cite[Proof of Theorem 1.2]{penrose} and \cite[Proof of Proposition 6.1]{penrose2016connectivity}.
A wireless network with $n$ nodes is often modeled by a \emph{random
geometric graph}~\cite{penrose,penrose2016connectivity}
 $G_{RGG}(n, r_n,\mathcal{A})$, where $n$ nodes
  are uniformly and independently distributed in a network field $\mathcal{A}$
   and two nodes have an edge in between  if and only if their distance is at most the transmission range $r_n$. ($k$-)Connectivity in $G_{RGG}(n, r_n,\mathcal{A})$ has been widely
investigated in the literature~\cite{penrose,penrose2016connectivity,Gupta98criticalpower,LiWanWangYi,wanAsymptoticCritical,ta2009phase}, where $\mathcal{A}$ may exhibit the boundary effect and letting $\mathcal{A}$ be a torus eliminates the boundary effect~\cite{mao2013connectivity,JZNodeCapture}. Gupta and Kumar \cite{Gupta98criticalpower}
   show that with $\mathcal{D}$
  being a disk of unit area,
  graph
  $G_{RGG}(n, r_n,\mathcal{D})$ is {\textit{a.a.s.}} connected if and only if the sequence $\alpha_n$ defined by $\pi {r_n}^2 = \frac{\ln n + \alpha_n}{n}$ satisfies
  $\lim_{n\to\infty}\alpha_n= \infty$. Penrose \cite{penrose} extend the result to $k$-connectivity for
$G_{RGG}(n, r_n,\mathcal{T})$ on a torus $\mathcal{T}$.
Penrose \cite{penrose} also studies
$k$-connectivity in graph $G_{RGG}(n, r_n,\mathcal{S})$ on the square
$\mathcal{S}$, while the exact formula of $r_n$ to ensure $k$-connectivity is obtained later by Li~\emph{et~al.}~\cite{LiWanWangYi} as well as by Wan and Yi
\cite{wanAsymptoticCritical}. To further characterize the $k$-connectivity behavior,
Ta~\emph{et~al}.~\cite{ta2009phase} derive the phase transition width of $k$-connectivity in a $d$-dimensional random geometric graph for $d=1,2,3$.


The disk model has been generalized to represent more generic wireless connections. One generalization called the \textit{general connection model} has received much interest \cite{mao2013connectivity,mao2011asymptotic,ta2009connectivity,mao2012towards}. In this model, two nodes separated by
a distance $x$ are directly connected with probability
$f(x)$ for a function $f:[0,\infty) \to [0,1]$, independent of the event that any other pair of
nodes are directly connected. Mao and Anderson~\cite{mao2013connectivity,mao2011asymptotic} obtain a strong connectivity result of wireless networks under this general connection model and under the Poisson node distribution, where the nodes are distributed according to a Poisson point process. Their connectivity result under the general connection model generalizes the result under the (traditional) disk model by Gupta and Kumar \cite{Gupta98criticalpower} for the disk model. An early analysis of connectivity of wireless networks under the general connection model is presented by Ta~\emph{et~al}.~\cite{ta2009connectivity}, where they prove the
probability of connectivity is asymptotically equivalent to the probability
of having no isolated node. By analyzing the number of isolated nodes under the general connection model, Mao and Anderson~\cite{mao2012towards} show the differences between the dense network
model, the extended network model, and  the infinite network
model. For a comprehensive discussion of connectivity in wireless networks under the general connection model or other alternatives, we refer interested readers to an excellent book by Mao~\cite{mao2017connectivity}.


\textbf{Connectivity of wireless networks under the log-normal connection model.} Despite being very useful, the above general connection model and its special case, the disk model, have a major limitation: connections are assumed to be independent in some sense; more specifically, as long as two nodes are within certain distance, they have a link in between (or with some probability in the general connection model), no matter how many other communicating nodes are nearby. The above assumption may not hold in reality due to the interference between connections. Taking into account of this, the following log-normal connection model has been considered~\cite{hekmat2006connectivity,yang2012connectivity}. In this model, two nodes are
directly connected if the received power at one node from the
other node, whose attenuation obeys the log-normal model, is at least a given threshold~\cite{yang2012connectivity}. Hekmat and Van Mieghem~\cite{hekmat2006connectivity} investigate connectivity  under the log-normal connection model, but their results assume that the node isolation events are
independent. Without relying on this assumption,
Yang~\emph{et~al}.~\cite{yang2012connectivity} provide more rigorous results by showing a necessary condition and a sufficient condition to ensure connectivity, where the bounds in  the two parts differ by a constant factor only.

\textbf{Connectivity of wireless information-theoretic secure networks}. In addition to the use of cryptographic techniques, security of wireless networks has also been studied from the information-theoretic perspective, where physical layer techniques are utilized to protect communications~\cite{haenggi2008secrecy,yang2014connectivity}. This thread of research is orthogonal to our work.

\textbf{Connectivity of wireless networks under the disk model with unreliable links.} A wireless network under the disk model on a network area $\mathcal{A}$ with unreliable links can be modeled by the intersection of a random geometric graph $G_{RGG}(n, r_n,\mathcal{A})$ and an Erd\H{o}s--R\'{e}nyi graph $G(n,p_n)$.  Below we discuss studies of this graph intersection in the literature.

 For graph $G(n,p_n) \bcap
G_{RGG}(n, r_n,\mathcal{A})$, Yi~\emph{et~al.}~\cite{Yi06GLOBECOM} investigate the distribution for the number
of isolated nodes. Yi~\emph{et~al.}~\cite{Yi06GLOBECOM,yi2006asymptotic} also explore the impact of unreliable nodes to the number
of isolated nodes.  Gupta and Kumar~\cite{Gupta98criticalpower} present connectivity results for random geometric graph $G_{RGG}(n, r_n,\mathcal{A})$. In the same work~\cite{Gupta98criticalpower}, they also propose the Gupta--Kumar
  conjecture for connectivity in the intersection of a random geometric graph and an Erd\H{o}s--R\'{e}nyi graph. Specifically, the conjecture states that under
  $\pi  {r_n}^2 p_n = \frac{\ln n + \alpha_n}{n}$,
  graph $G_{RGG}(n, r_n,\mathcal{D}) \bcap G(n,p_n)$
on the disk $\mathcal{D}$ of unit area is \textit{a.a.s.} connected if
and only if
  $\lim_{n\to \infty}\alpha_n= \infty$. One significant attempt to answer the Gupta--Kumar conjecture is the work by Pishro-Nik~\textit{et al.}~\cite{Pishro}, where $G_{RGG}(n, r_n,\mathcal{S}) \bcap G(n,p_n)$ on a unit square $\mathcal{S}$ is considered. Yet,  they assume that $G_{RGG}(n, r_n,\mathcal{S}) \bcap G(n,p_n)$ is $k$-connected whenever its minimum degree is at least $k$. This assumption is verified by Penrose~\cite{penrose2016connectivity} recently with a lengthy proof. In fact, the results of Penrose~\cite{penrose2016connectivity} also address the Gupta--Kumar conjecture. The difficulty of the conjecture is to analyze the connection structure when two distinct kinds of   graphs intersect: even if individual graphs are highly connected, the resulting topology after intersection can still become disconnected. Penrose~\cite{penrose2016connectivity} obtain that the connectivity result of $G_{RGG}(n, r_n,\mathcal{T}) \bcap G(n,p_n)$ on a unit torus $\mathcal{T}$ resembles the Gupta--Kumar
  conjecture, but the connectivity result of $G_{RGG}(n, r_n,\mathcal{S}) \bcap G(n,p_n)$ on a unit square $\mathcal{S}$ is more complex. According to Penrose~\cite{penrose2016connectivity}, the underlying reason for different connectivity results under the torus and under the square is the impact of the boundary effect on
 the asymptotics for the number of isolated nodes.\label{relatedend}

\section{Ideas for Proving Theorem \ref{thm:OneLaw+NodeIsolation}}
\label{sec:ProofTheoremNodeIsolation}


In this section, we explain  the basic ideas to prove Theorem~\ref{thm:OneLaw+NodeIsolation}. We first introduce an additional condition $|{\alpha_n} |=  o ( \ln n)$, and then use the relationship between connectivity and the absence of isolated nodes.




We first show that
the extra condition $|{\alpha_n} |=  o ( \ln n)$
can be introduced in proving
Theorem~\ref{thm:OneLaw+NodeIsolation}, where $|{\alpha_n} |$ is the absolute value of $\alpha_n$. From (\ref{eq:scalinglaw}) in Theorem \ref{thm:OneLaw+NodeIsolation}, since ${\alpha_n}$ measures the deviation of the edge probability $t_n$
from the critical scaling $\frac{ \ln n + (k-1)\ln \ln n}{n}$, we call the extra condition $|{\alpha_n} |=  o ( \ln n)$ as \emph{the confined deviation}.
Then our goal   is to show
\begin{align}
\text{Theorem \ref{thm:OneLaw+NodeIsolation} with the confined deviation} ~~\Longrightarrow ~~
\text{Theorem \ref{thm:OneLaw+NodeIsolation}}.
\label{with_extra}
\end{align}

We write $t_n$ back as
$t (K_n, P_n, q, {p_n})$ and remember that given
$K_n$, $P_n$, $q$ and ${p_n}$, one can determine
$\alpha_n$ from (\ref{u4}) and (\ref{eq:scalinglaw}).
 To show (\ref{with_extra}), we first present Lemma \ref{lem_Gq_cplinga} on graph coupling~\cite{2013arXiv13010466R}.

\begin{lem} \label{lem_Gq_cplinga}
 {
For a graph $\mathbb{G}_q(n, K_n,P_n, {p_n})$ under $ P_n =
\Omega(n)$ and $\frac{{K_n}^2}{P_n} = o(1) $, with a sequence $\alpha_n$ defined by (\ref{eq:scalinglaw}) (i.e., $t_n  = \frac{\ln  n + (k-1)\ln \ln n   +
 {\alpha_n}}{n}$),
the following results hold:
\begin{itemize}
\item[(a)]
If $\lim_{n \to \infty}\alpha_n = -\infty$, there exists a graph $\mathbb{G}_q(n,
\widetilde{K_n},\widetilde{P_n}, \widetilde{{p_n}})$ under $\widetilde{P_n} = \Omega(n)$, $\frac{{\widetilde{K_n}}^2}{\widetilde{P_n}} = o(1) $ and
$ t(\widetilde{K_n},\widetilde{P_n}, q,\widetilde{p_n})  =  \frac{\ln  n + (k-1)\ln \ln n   + {\widetilde{\alpha_n}}}{n}$ with $\lim_{n \to \infty}\widetilde{\alpha_n} = -\infty$ and $\widetilde{\alpha_n} = -o(\ln n)$,
such that there exists a graph coupling under which
$\mathbb{G}_q(n, K_n,P_n, {p_n})$ is a spanning subgraph of $\mathbb{G}_q(n,\widetilde{K_n},\widetilde{P_n}, \widetilde{{p_n}})$.
\item[(b)] If $\lim_{n \to \infty}\alpha_n = \infty$, there exists a graph $\mathbb{G}_q(n,\widehat{K_n},\widehat{P_n}, \widehat{{p_n}})$ under $\widehat{P_n} = \Omega(n)$, $\frac{{\widehat{K_n}}^2}{\widehat{P_n}} = o(1) $ and
$ t(\widehat{K_n},\widehat{P_n}, q,\widehat{p_n})   = \frac{\ln  n+ (k-1)\ln \ln n   + {\widehat{\alpha_n}}}{n}$
with $\lim_{n \to \infty}\widehat{\alpha_n} = \infty$ and $\widehat{\alpha_n} = o(\ln n)$,
such that there exists a graph coupling under which
$\mathbb{G}_q(n, K_n,P_n, {p_n})$ is a spanning supergraph of $\mathbb{G}_q(n,\widehat{K_n},\widehat{P_n}, \widehat{{p_n}})$.
\end{itemize}
 }

\end{lem}

For any graph that is not $k$-connected, its spanning subgraph is not $k$-connected. Also, for any $k$-connected graph, its spanning supergraph is $k$-connected. Given the above, Lemma~\ref{lem_Gq_cplinga} clearly implies (\ref{with_extra}).
 {Hence, in proving \mbox{Theorem \ref{thm:OneLaw+NodeIsolation}}, we can always assume the confined deviation $|{\alpha_n} |=  o ( \ln n)$.} In the rest of the paper, {we often write $\mathbb{G}_q(n,
K_n, P_n, {p_n})$ as $\mathbb{G}_q$} for notation brevity.

Given the conditions of Theorem \ref{thm:OneLaw+NodeIsolation} (i.e., $ P_n =
\Omega(n)$ and $\frac{{K_n}^2}{P_n} = o(1)$), and the extra $|\alpha_n| = o(\ln n) $ introduced  in Section~\ref{sec:ProofTheoremNodeIsolation}, we utilize Lemma \ref{lemboundKn} to have $K_n = \Omega\Big( n^{\frac{q-1}{2q}}
(\ln n )^{\frac{1}{2q}} \Big) = \omega(1)$. Then given $K_n  = \omega(1)$ and $\frac{{K_n}^2}{P_n} = o(1)$, we use
\cite[Theorem 1]{QcompTech14} to obtain
 \begin{align}
&  \lim_{n \rightarrow \infty } \mathbb{P}\bigg[
\begin{array}{c}
\mathbb{G}_q~\mbox{has a minimum} \\
\mbox{node degree at least $k$.}
\end{array}
\bigg] \nonumber
\end{align}
\begin{subnumcases}{=}  0,\quad\text{if  }\lim_{n \to \infty} \alpha_n
=- \infty, \label{thm-mnd-eq-0}  \\
1,\quad\text{if  }\lim_{n \to \infty} \alpha_n
= \infty. \label{thm-mnd-eq-1}
\end{subnumcases}
Since a necessary condition for a graph to be $k$-connected is that
the minimum node degree is at least $k$, (\ref{thm-mnd-eq-0}) clearly implies the zero-law (\ref{thm-con-eq-0}) of $k$-connectivity. Moreover, given (\ref{thm-mnd-eq-1}), the one-law
 (\ref{thm-con-eq-1}) of $k$-connectivity will be proved once we show Lemma \ref{lem_Gq_no_isolated_but_not_conn} below. Note that we can introduce   $|\alpha_n| = o(\ln n) $ from the argument in Section~\ref{sec:ProofTheoremNodeIsolation}.

\begin{lem} \label{lem_Gq_no_isolated_but_not_conn}

{
For a graph $\mathbb{G}_q(n, K_n,P_n, {p_n})$ under $ P_n =
\Omega(n)$ and $\frac{{K_n}^2}{P_n} = o(1)$, if the sequence $\alpha_n $ defined by (\ref{eq:scalinglaw}) satisfies
$\lim_{n \to \infty}\alpha_n = \infty$ and $|{\alpha_n} |=  o ( \ln n)$, then
\begin{align}
\lim_{n \to \infty}
 \mathbb{P} \left[ \begin{array}{l}
 \mathbb{G}_q\mbox{ has a minimum node degree at least $k$},\\ \mbox{but is not $k$-connected.} \end{array} \right]
 = 0.  \label{eq:OneLawAfterReductionsb}
 \end{align}
 }
\end{lem}

Lemma \ref{lem_Gq_no_isolated_but_not_conn} is established in Section \ref{sec:lem_Gq_no_isolated_but_not_conn}.
Due to space limitation, we provide many details in the full version \cite{fullversion}.

\section{Establishing Lemma \ref{lem_Gq_no_isolated_but_not_conn}}
\label{sec:lem_Gq_no_isolated_but_not_conn}

For a graph, let its \textit{node connectivity} be the minimum number of nodes that need to be removed to disconnect the remaining nodes from each other. Then a graph is $k$-connected if and only if its node connectivity is at least $k$. A graph is not $k$-connected if and only if its node connectivity is less than $k$. To prove Lemma \ref{lem_Gq_no_isolated_but_not_conn}, we have
\begin{align}
&  \mathbb{P} \left[\begin{array}{l}
\mathbb{G}_q\mbox{ has a minimum node degree at least $k$,} \\ \mbox{but is not $k$-connected.} \end{array}\right]
\nonumber \\ &   \leq \sum_{\ell=0}^{k-1} \mathbb{P}\left[\begin{array}{l} \mbox{$\mathbb{G}_q$'s node connectivity equals $\ell$, and}\\ \mbox{$\mathbb{G}_q$'s minimum node degree is greater than $\ell$} \end{array}\right]. \label{Fsumell}
 \end{align}

We define event $F_{n,\ell}$ as follows:
\begin{align}
 F_{n,\ell}:&~\text{the event that $\mathbb{G}_q$'s node connectivity equals $\ell$,}  \nonumber \\ & ~\text{and $\mathbb{G}_q$'s minimum node degree is greater than $\ell$.} \label{Fsumelldefine}
 \end{align}
 Then the summation in (\ref{Fsumell}) becomes $\sum_{\ell=0}^{k-1} \bP{F_{n,\ell}}$.
 The   idea~\cite{ZhaoYaganGligor} in establishing Lemma \ref{lem_Gq_no_isolated_but_not_conn} is to find an
upper bound on $\sum_{\ell=0}^{k-1} \bP{F_{n,\ell}}$
and   show that this bound goes to zero as $n\to \infty$.


We begin by finding the needed upper bound. Let $\mathcal{N}$ denote the collection of all non-empty
subsets of the node set $\{ v_1, \ldots , v_n \}$ in graph $\mathbb{G}_q$.
Recalling that $S_i$ denotes  the set of $K_n$ keys on node
$v_i $,
we introduce an
event $E_n(\boldsymbol{X}_n)$ in the following manner:
\begin{equation}\nonumber
E_n(\boldsymbol{X}_n)= \bigcup_{T \subseteq \mathcal{N}: ~
|T| \geq 1} ~ \left[\left|\cup_{j \in T}
S_j\right|~\leq~{X}_{n,|T|}\right]
\label{eq:E_n_defn}
\end{equation}
where
$\boldsymbol{X}_n=[{X}_{n,1},~{X}_{n,2},~
\ldots,~ {X}_{n,n}]$ is an $n$-dimensional integer-valued
array. We define $r_n^{*}$ by
\begin{eqnarray}
r_n^{*}  := \min \left ( \left
\lfloor \frac{P_n}{K_n} \right \rfloor, ~\left \lfloor \frac{n }{2}
\right \rfloor \right ) .  \label{eq:rstardef}
\end{eqnarray}
We set
\begin{align}
 X_{n,i}   =
\begin{cases}
K_n, &   \text{ for }i=1,
\\
\max\{ \left \lfloor (1+\varepsilon) K_n \right \rfloor,
\left \lfloor \lambda K_n i \right \rfloor \}, &\text{ for }i=2,\ldots,  r_n^{*},\\
 \left \lfloor\mu P_n \right \rfloor, &    \hspace{-15pt}  \text{for }i= r_n^{*}+1, \ldots, n,
\end{cases} \label{eq:X_S_theta}
\end{align}
for an arbitrary constant $0<\varepsilon<1$ and constants $\lambda$  and $\mu$ specified below. Recalling the condition $P_n = \Omega(n)$, we let $P_n \geq \sigma n$ for all $n$ sufficiently large, where $\sigma$ is certain positive constant. We select $\lambda$  and $\mu$ satisfying $0<\lambda<\frac{1}{2}$, $\max \left( 2 \lambda \sigma , \lambda \left( \frac{e^2}{\sigma}
\right) ^{\frac{ \lambda }{ 1 - 2 \lambda } } \right) < 1$, $0<\mu<\frac{1}{2}$  and $\max \left ( 2 \left ( \sqrt{\mu} \left ( \frac{e}{ \mu } \right
)^{\mu} \right )^\sigma, \sqrt{\mu} \left ( \frac{e}{ \mu }
\right)^{\mu} \right ) < 1$, such that the event $E_n(\boldsymbol{X}_n)$ defined above satisfies
\begin{eqnarray}
\lim_{n \rightarrow \infty} \bP{E_n(\boldsymbol{X}_n)} =
0. \label{eq:X_S_thetalim0}
\end{eqnarray}

Given
$\bP{F_{n,\ell}}
 \leq
\bP{E_n(\boldsymbol{X}_n)} + \bP{ F_{n,\ell} \cap \overline{E_n(\boldsymbol{X}_n)} }$,
and (\ref{eq:X_S_thetalim0}), we will obtain the result $\lim_{n \to \infty}\bP{ F_{n,\ell} }=0$
once establishing the following proposition. After   showing $\lim_{n \to \infty}\bP{ F_{n,\ell} }=0$, since $k$ does not scale with
$n$, we further derive  $\lim_{n \to \infty} (\sum_{\ell=0}^{k-1} \bP{F_{n,\ell}})=0$, which along with (\ref{Fsumell}) and (\ref{Fsumelldefine}) completes proving Lemma \ref{lem_Gq_no_isolated_but_not_conn}.

\begin{proposition}
For a graph $\mathbb{G}_q(n, K_n,P_n, {p_n})$ under $ P_n =
\Omega(n)$ and $\frac{{K_n}^2}{P_n} = o(1)$, if the sequence $\alpha_n $ defined by (\ref{eq:scalinglaw}) (i.e., $t_n  = \frac{\ln  n + (k-1)\ln \ln n  +
 {\alpha_n}}{n}$) satisfies
$\lim_{n \to \infty}\alpha_n = \infty$  and $|{\alpha_n} |=  o ( \ln n)$, then for $\ell = 0,1,\ldots,k-1$, we have $\lim_{n \to \infty}\bP{ F_{n,\ell} \cap \overline{E_n(\boldsymbol{X}_n)} } = 0.$\label{prop:OneLawAfterReductionPart2}
\end{proposition}

\noindent \textbf{Proof of Proposition \ref{prop:OneLawAfterReductionPart2}:}

Recall that the node set of graph $\mathbb{G}_q$ is $\mathcal{V}_n = \{v_1, v_2,\ldots,v_n\}$, and $F_{n,\ell}$ denotes the event that graph $\mathbb{G}_q$'s node connectivity equals $\ell$, and $\mathbb{G}_q$'s minimum node degree is greater than $\ell$. Below we analyze the graph structure of $\mathbb{G}_q$ when event $F_{n,\ell}$ happens. When graph $\mathbb{G}_q$'s node connectivity equals $\ell$, we have by definition that there exists a subset $U$ of the node set $\mathcal{V}_n = \{v_1, v_2,\ldots,v_n\}$ nodes with  $|U| = \ell$ such that $\mathbb{G}_q(\mathcal{V}_n \setminus U)$ is disconnected, where $\mathbb{G}_q(\mathcal{V}_n \setminus U)$ denotes the subgraph of $\mathbb{G}_q$ with the node set restricted to $\mathcal{V}_n \setminus U$. We consider $n \geq \ell + 3$ so $\mathbb{G}_q(\mathcal{V}_n \setminus U)$ has at least three nodes. Since $\mathbb{G}_q(\mathcal{V}_n \setminus U)$ is disconnected,
$\mathbb{G}_q(\mathcal{V}_n \setminus U)$   has a set of components (say $m$ components where $m\geq 2$) such that  the following $\raisebox{.5pt}{\textcircled{\raisebox{-.9pt} {a}}} $ and $\raisebox{.5pt}{\textcircled{\raisebox{-.9pt} {b}}} $ both happen: $\raisebox{.5pt}{\textcircled{\raisebox{-.9pt} {a}}} $ each component is either self-connected or has only one node; $\raisebox{.5pt}{\textcircled{\raisebox{-.9pt} {b}}} $ different components are disconnected from each other. Considering that $\mathbb{G}_q(\mathcal{V}_n \setminus U)$ has $m$ components in total for some $m\geq 2$, given $   |\mathcal{V}_n \setminus U| = n - \ell$, we  pick one component with at most  $ \lfloor \frac{n - \ell}{2} \rfloor $ nodes, and call this component $S$. Below we explain that $S$ cannot have  only one node. By contradiction, if $S$ has  only one node, supposing that this node is $v_*$, then $v_*$ does not have neighbors in $\mathcal{V}_n \setminus U$, meaning that $v_*$'s neighbors in $\mathbb{G}_q$ all belong to the set $U$. Hence, with $|U| = \ell$, $v_*$'s degree in $\mathbb{G}_q$ is at most $\ell$,  contradicting with the condition that $\mathbb{G}_q$'s minimum  degree is greater than $\ell$.
 Summarizing the above analysis, whenever $F_{n,\ell}$ happens, there
exist disjoint subsets $U,S$ of the node set $\mathcal{V}_n = \{v_1, v_2,\ldots,v_n\}$  with $|U| = \ell$ and $2 \leq |S| \leq  \lfloor \frac{n - \ell}{2} \rfloor$ such that
\begin{itemize}
  \item[\ding{172}] with  $\mathbb{G}_q(S)$ denoting the subgraph of $\mathbb{G}_q$ with the node set restricted to $S$, $\mathbb{G}_q(S)$ is connected;
  \item[\ding{173}] with  $\mathbb{G}_q(\mathcal{V}_n \setminus U)$ denoting the subgraph of $\mathbb{G}_q$ with the node set restricted to $\mathcal{V}_n \setminus U$, $S$ is isolated in
$\mathbb{G}_q(\mathcal{V}_n \setminus U)$.
\end{itemize}

We further analyze the graph structure of $\mathbb{G}_q$ when event $F_{n,\ell}$ happens. We let $v_{\#} $ be an arbitrary node in set $U$ (recall $|U| = \ell$). Since graph $\mathbb{G}_q$'s node connectivity equals $\ell$ under $F_{n,\ell}$, deleting the $\ell-1$ nodes of $U\setminus \{v_{\#}\}$ in  $\mathbb{G}_q$ will still preserve connectivity of the remaining graph $\mathbb{G}_q ((\mathcal{V}_n \setminus U)\bcup \{v_{\#}\})$. Since we know from \ding{173} above that there is no edge  between any node in $S$ and any node in $(\mathcal{V}_n \setminus U)\setminus S$, to ensure connectivity of   $\mathbb{G}_q ((\mathcal{V}_n \setminus U)\bcup \{v_{\#}\})$,   we have
\begin{itemize}
  \item[\ding{174}] for any node $v_{\#} $ in set $U$, $v_{\#} $ has at least one neighbor in $S$ {and} at least one neighbor in $(\mathcal{V}_n \setminus U)\setminus S$.
\end{itemize}


Now we define  $C_n(S)$ and $D_n(S, U)$ to represent \ding{172} and \ding{173} above. In addition, we  define $B_n(S, U)$ as the event that any node in set $U$ has at least one neighbor in $S$; i.e., $B_n(S, U)$ relaxes the requirement in \ding{174} above. \label{defineeventBCD} Summarizing the above, we know that $F_{n,\ell}$ is a subevent of $\bigcup_{ \begin{subarray}{l}
|U| = \ell, \\ 2 \leq |S| \leq \lfloor
\frac{n-\ell }{2} \rfloor
\end{subarray} } [B_n(S, U) \cap C_n(S) \cap D_n(S, U)] $. We let $\mathcal{N}_{n,\ell} $ be the collection of the subsets of  $\mathcal{V}_n$ with exactly $\ell$ elements, and let $\mathcal{N}_{r}(\mathcal{V}_n \setminus U) $ be the collection of the subsets of $\mathcal{V}_n \setminus U$ with exactly $r$ elements. Then from the union bound, we obtain
\begin{align}\nonumber
 \nonumber
&\bP{ F_{n,\ell} \cap
\overline{E_n(\boldsymbol{X}_n)}} \nonumber \\&\leq    \sum_{U \in
\mathcal{N}_{n,\ell}} \sum_{r=2}^{ \lfloor
\frac{n-\ell }{2} \rfloor}   \sum_{S \in
\mathcal{N}_{r}(\mathcal{V}_n \setminus U) } \nonumber \\& ~~~~~\bP{B_n(S, U) \cap C_n(S) \cap D_n(S, U) \cap
\overline{E_n(\boldsymbol{X}_n)}}.  \label{pfnboundnew}
\end{align}

For each $r=2,3,\ldots, \lfloor
\frac{n-\ell }{2} \rfloor$, when $S$ is $\{ v_1, \ldots , v_r \}$, and $U$ is $\{ v_{r+1}, \ldots , v_{r+\ell} \}$, we let $B_n(S, U) ,  C_n(S) , D_n(S, U)$ be $B_{n,r,\ell},C_{n,r},D_{n,r,\ell}$.   We further define
 $A_{n,r,\ell}:= B_{n,r,\ell} \bcap C_{n,r} \bcap D_{n,r,\ell} .$
Then by exchangeability, we obtain from (\ref{pfnboundnew}) that
%
%
\begin{align}
& \bP{F_{n,\ell}\cap
\overline{E_n(\boldsymbol{X}_n)}} \nonumber \\ &   \leq \sum_{r=2}^{ \lfloor
\frac{n-\ell }{2} \rfloor }  {n \choose \ell}  {n - \ell \choose r} ~ \bP{ A_{n,r,\ell} \cap
\overline{E_n(\boldsymbol{X}_n)}},
\label{eq:BasicIdea+UnionBound2}
\end{align}
where we use $
|\mathcal{N}_{n,\ell}| = {n \choose \ell} $ and $|\mathcal{N}_{r}(\mathcal{V}_n \setminus U)| = {n - \ell \choose r} $.
Then the proof of Proposition
\ref{prop:OneLawAfterReductionPart2} will be completed once we show
\begin{align}
 & \lim_{n \rightarrow \infty} \sum_{r=2}^{ \lfloor \frac{n-\ell }{2}
\rfloor } {n \choose \ell} {n-\ell \choose r}  ~ \bP{ A_{n,r,\ell} \cap
\overline{E_n(\boldsymbol{X}_n)}} = 0,\nonumber \\  &   \text{for $\ell = 0,1,\ldots,k-1$}. \label{eq:OneLawToShow}
\end{align}

We now analyze $A_{n,r,\ell}:= B_{n,r,\ell} \bcap C_{n,r} \bcap D_{n,r,\ell}$.  For each $j=r+1,\ldots,n$, we define $u_{r,j}$ as the set of nodes, each of which belongs to $\{v_1,\ldots,v_r\}$ and also has an ``on'' channel with node $v_j$. For
 $j=r+1, \ldots
r+\ell $, we define
\begin{eqnarray}
 \mathcal{B}_{n,r,\ell}^{(j)} : =  \cup_{i \in u_{r,j}} \Gamma_{ij} ,
\label{probBeve}
\end{eqnarray}
and
 for   $j=r+\ell+1,\ldots,n$, we define
\begin{eqnarray}
\mathcal{D}_{n,r,\ell}^{(j)}:=\cap_{i \in u_{r,j}} \overline{\Gamma_{ij}}.
\label{mathcalDnrjq}
\end{eqnarray}
Then we have
\begin{align}
 B_{n,r,\ell} & = \bigcap_{ j=r+1}^{r+\ell} \mathcal{B}_{n,r,\ell}^{(j)},
  \text{ and } D_{n,r,\ell} & = \bigcap_{ j=r+\ell+1}^n \mathcal{D}_{n,r,\ell}^{(j)}.
\label{Dgem}
\end{align}


  Conditioning on the random variables $\{S_i, \ i=1, \ldots , r\} $ and $\{ \boldsymbol{1}[L_{ij}], \ i,j=1,\ldots, r \}$
(these two sets   determine the event $C_{n,r}$),
 the events \mbox{$\{ \mathcal{B}_{n,r,\ell}^{(j)},~j=r+1, \ldots
,r+\ell\}$}  and $\{ \mathcal{D}_{n,r,\ell}^{(j)},~j=r+\ell+1, \ldots
,n\}$ are all conditionally independent. Then we conclude via $A_{n,r,\ell}:= B_{n,r,\ell} \bcap C_{n,r} \bcap D_{n,r,\ell}$ and (\ref{Dgem}) that
  \begin{align}
 &   \bP{ A_{n,r,\ell} \bcap
\overline{E_n(\boldsymbol{X}_n)}}    \nonumber \\
 &    =   \mathbb{E}\scalebox{1.7}{\Bigg[}\1{ C_{n,r}
 \bcap
\overline{E_n(\boldsymbol{X}_n)}
 }  \nonumber \\
 &   ~~~~~  \times   \prod_{ j=r+1}^{r+\ell} \bP{ \mathcal{B}_{n,r,\ell}^{(j)}~\Bigg |\begin{array}{r}
  S_i, ~~~~~\ i=1, \ldots , r, \\ \boldsymbol{1}[L_{ij}], ~~~~~\ i=1,\ldots, r, \\ j=r+1, \ldots
r+\ell .
\end{array}} \nonumber \\
 &   ~~~~~  \times   \prod_{j=r+\ell+1}^n \bP{ \mathcal{D}_{n,r,\ell}^{(j)}~\Bigg |\begin{array}{r}
  S_i, ~~~~~\ i=1, \ldots , r, \\ \boldsymbol{1}[L_{ij}], ~~~~~\ i=1,\ldots, r, \\ j=r+\ell+1, \ldots
n .
\end{array}}\scalebox{1.7}{\Bigg]}, \label{boundDrleqab}
\end{align}
where the expectation is taken over random variables $\{S_i, \ i=1, \ldots , r\} $ and $\left\{  \boldsymbol{1}[L_{ij}], \begin{array}{r}
  i=1, \ldots , r,  \\  j=r+1, \ldots
n .
\end{array}  \right\}$.

For $j=r+1,\ldots,r+\ell $, from (\ref{probBeve}),
it holds by the union bound that
\begin{align}
 &  \bP{ \mathcal{B}_{n,r,\ell}^{(j)}~\Bigg |\begin{array}{r}
  S_i, ~~~~~\ i=1, \ldots , r, \\ \boldsymbol{1}[{L}_{ij}], ~~~~~\ i=1,\ldots, r, \\ j=r+1, \ldots
r+\ell .
\end{array}} \nonumber \\
 &   \leq \sum_{i \in u_{r,j}} \mathbb{P} \big[  \Gamma_{ij} ~|~   S_i \big] = \sum_{i \in u_{r,j}} s_n
 = s_n |u_{r,j}| . \label{boundBrleq}
\end{align}
With $u_{r,j} = \sum_{i=1}^{r} \boldsymbol{1}[L_{ij}]$,     $|u_{r,j}|$ follows a binomial distribution with $r$ trials and the success probability ${p_n}$ in each trial. Hence, from $t_n = s_n p_n$, it holds that
  \begin{align}
   \mathbb{E}\left[ s_n |u_{r,j}|\right]  = s_n \cdot rp_n = r t_n. \label{ecnrbp2}
  \end{align}

 Given $\{S_i, \ i=1, \ldots , r\} $ and $\{ \boldsymbol{1}[{L}_{ij}], \ i,j=1,\ldots, r \}$, the probability of $\big[
|(\bigcup_{i \in u_{r,j}} S_i)  \cap S_j| \geq q \big]$ is given by $\frac{\binom{|\bigcup_{i \in u_{r,j}} S_i|}{q}\binom{K_n}{q}}{\binom{P_n}{q}}$. \vspace{2pt} Then on the event $\overline{E_n(\boldsymbol{X}_n)}$ in (\ref{eq:E_n_defn}) which ensures $|\cup_{i \in u_{r,j}} S_i|> {X}_{n,|u_{r,j}|} $,  it follows that
 \begin{align}
&  \bP{ \mathcal{D}_{n,r,\ell}^{(j)}~\Bigg |\begin{array}{r}
  S_i, ~~~~~\ i=1, \ldots , r, \\ \boldsymbol{1}[{L}_{ij}], ~~~~~\ i=1,\ldots, r, \\ j=r+\ell+1, \ldots
n .
\end{array}}  \leq 1-\frac{\binom{X_{|u_{r,j}|,n}}{q}\binom{K_n}{q}}{\binom{P_n}{q}}.
   \label{bnmbcpinusa}
\end{align}

 Below we we  will prove that on the event $\overline{E_n(\boldsymbol{X}_n)}$, it holds
for all $n$ sufficiently large that
  \begin{align}
 \mathbb{E}\scalebox{1.25}{\Bigg[} 1-\frac{\binom{X_{|u_{r,j}|,n}}{q}\binom{K_n}{q}}{\binom{P_n}{q}} \scalebox{1.25}{\Bigg]}  \leq g_{r,n},
   \label{pnsnrlabcrt2}
\end{align}
 for function $g_{r,n}$ defined by \begin{align}
g_{r,n}: =
  \begin{cases}
  \min\big\{e^{-(1+\frac{\varepsilon_2}{2}) t_n},e^{-\lambda_2 t_n r} \big\}, ~\text{for } r=2,\ldots,r_n^*, \\
  e^{-\lambda_2 t_n r} + e^{- \mu_2 K_n}, ~\text{for } r=r_n^*+1,\ldots,n.
  \end{cases} \label{defgnr}
\end{align}


In view of (\ref{boundDrleqab})--(\ref{pnsnrlabcrt2}), considering the mutual independence
among $\big\{|u_{r,j}|\big\}\big|_{j=r+1,\ldots,n}$ and
$\1{ C_{n,r}
 \bcap
\overline{E_n(\boldsymbol{X}_n)}
 } $,
 and
using $ \mathbb{E}\big[\boldsymbol{1}[{ C_{n,r}  \bcap
\overline{E_n(\boldsymbol{X}_n)} }]\big]  \leq   \mathbb{P}[{ C_{n,r} }] $,   we obtain
 \begin{align}
& \bP{ A_{n,r} \bcap
\overline{E_n(\boldsymbol{X}_n)}}
\nonumber \\
 &   \leq   \mathbb{P}[{ C_{n,r} }]  \times \min\{ (r t_n )^{\ell},~1\}
\times {g_{r,n}}^{n-r-\ell} .  \label{ecnrbp1}
\end{align}
%
%
%

To establish (\ref{pnsnrlabcrt2}), below we we  first prove that on the event $\overline{E_n(\boldsymbol{X}_n)}$, it holds
for all $n$ sufficiently large that
\begin{align}
& 1-\frac{\binom{X_{|u_{r,j}|,n}}{q}\binom{K_n}{q}}{\binom{P_n}{q}}  \leq f(|u_{r,j}|) \label{olp_xjdefead2}\end{align} for $f(|u_{r,j}|)$ defined by \begin{align}
 & f(|u_{r,j}|): = \nonumber \\
 &   \begin{cases}
1-s_n,  &   \text{ for }|u_{r,j}|=1,
\\
 (1- s_n)^{ \max\{ (1+\varepsilon_2) ,~ \lambda_2 |u_{r,j}| \}} , &   \text{ for }|u_{r,j}|=2,\ldots,  r_n^{*},\\
e^{- \mu_2 K_n} &  \text{ for }|u_{r,j}|= r_n^{*}+1, \ldots, n.
\end{cases} \label{olp_xjdefead2fdef}
\end{align}

We now   establish (\ref{olp_xjdefead2}) and   use (\ref{olp_xjdefead2}) to show (\ref{pnsnrlabcrt2}).
 We will use the following result given by \cite[Lemma 6]{bloznelis2013}:
 \begin{align}
s_n \leq \frac{\big[\binom{K_n}{q}\big]^2}{\binom{P_n}{q}}.   \label{bloznelis2013xa}
\end{align}

For $|u_{r,j}|=1$, it holds that
 \begin{align}
1 - \frac{\binom{X_{|u_{r,j}|,n}}{q}\binom{K_n}{q}}{\binom{P_n}{q}} & = 1 - \frac{\big[\binom{K_n}{q}\big]^2}{\binom{P_n}{q}} \leq
1- s_n.   \label{bnmbcpinu1veead1ab}
\end{align}

For $|u_{r,j}|=2,\ldots,r_n^*$, it holds that
 \begin{align}
  \frac{\binom{X_{|u_{r,j}|,n}}{q}\binom{K_n}{q}}{\binom{P_n}{q}}  & \geq s_n \cdot\frac{\binom{X_{|u_{r,j}|,n}}{q}}{\binom{K_n}{q}}
  \nonumber \\
 &    = s_n \cdot   \max\left\{\frac{\binom{\lfloor(1+\varepsilon) K_n\rfloor}{q}}{\binom{K_n}{q}},\frac{\binom{\lfloor \lambda K_n |u_{r,j}|\rfloor}{q}}{\binom{K_n}{q}}\right\}.  \label{bnmbcpinu2st}
\end{align}
From Lemma \ref{lemboundKn}-Property (ii), we obtain $K_n = \Omega\Big( n^{\frac{1}{2}-\frac{1}{2q}}(\ln n )^{\frac{1}{2q}} {p_n} ^{-\frac{1}{2q}} \Big)= \omega(1)$. As proved in the full version~\cite{fullversion},  given $K_n = \omega(1)$, for any constants $\varepsilon_2$ and $\lambda_2$ satisfying $0<\varepsilon_2 < (1+\varepsilon)^q-1$ and $ 0<\lambda_2 < {\lambda}^q < \big(\frac{1}{2}\big)^q < 1$, we have for all $n$ sufficiently large that
 \begin{align}
& \max\left\{\frac{\binom{\lfloor(1+\varepsilon) K_n\rfloor}{q}}{\binom{K_n}{q}},~\frac{\binom{\lfloor \lambda K_n |u_{r,j}|\rfloor}{q}}{\binom{K_n}{q}}\right\}  \nonumber \\
 &    \geq \max\{ (1+\varepsilon_2) , ~\lambda_2 |u_{r,j}| \}.   \label{bnmbcpinu1ve2}
\end{align}
 From $0\leq s_n \leq 1$ and $ \max\{ (1+\varepsilon_2) , \lambda_2 |u_{r,j}| \}  > 1$, we obtain
 \begin{align}
 &  1- s_n \cdot  \max\{ (1+\varepsilon_2) , \lambda_2 |u_{r,j}| \}
\nonumber \\
 &   \leq (1- s_n)^{ \max\{ (1+\varepsilon_2) , \lambda_2 |u_{r,j}| \}} .   \label{bnmbcpinu2}
\end{align}
Using (\ref{bnmbcpinu1ve2})  (\ref{bnmbcpinu2}) in (\ref{bnmbcpinu2st}), we have for $|u_{r,j}|=2,\ldots,r_n^*$ that
 \begin{align}
1 - \frac{\binom{X_{|u_{r,j}|,n}}{q}\binom{K_n}{q}}{\binom{P_n}{q}} &  \leq (1- s_n)^{ \max\{ (1+\varepsilon_2) , \lambda_2 |u_{r,j}| \}} .   \label{bnmbcpinu1veead1}
\end{align}

For $|u_{r,j}|=r_n^*+1,\ldots,n$, it holds that
 \begin{align}
1 - \frac{\binom{X_{|u_{r,j}|,n}}{q}\binom{K_n}{q}}{\binom{P_n}{q}} &  = 1 - \frac{\binom{\lfloor\mu P_n \rfloor}{q}\binom{K_n}{q}}{\binom{P_n}{q}} \leq e^{- \frac{\binom{\mu P_n}{q}\binom{K_n}{q}}{\binom{P_n}{q}} }.   \label{bnmbcpinu1veead1ax}
\end{align}
As proved by
 in the full version~\cite{fullversion}, for any constant $\mu_2$ satisfying $0<\mu_2 < (q!)^{-1}{\mu}^q$, we   obtain for all $n$ sufficiently large that
 \begin{align}
\frac{\binom{\lfloor\mu P_n\rfloor}{q}\binom{K_n}{q}}{\binom{P_n}{q}} & \geq \mu_2 K_n,   \label{bnmbcpinu1veead1ax2btep}
\end{align}
which with (\ref{bnmbcpinu1veead1ax}) further implies
 \begin{align}
1 - \frac{\binom{X_{|u_{r,j}|,n}}{q}\binom{K_n}{q}}{\binom{P_n}{q}} &  \leq e^{-\mu_2 K_n} \text{ for $|u_{r,j}|=r_n^*+1,\ldots,n$} .   \label{bnmbcpinu1veead1ax2b}
\end{align}

Summarizing (\ref{bnmbcpinu1veead1ab}) (\ref{bnmbcpinu1veead1}) and (\ref{bnmbcpinu1veead1ax2b}), on the event $\overline{E_n(\boldsymbol{X}_n)}$, we obtain (\ref{olp_xjdefead2})
for all $n$ sufficiently large.
Now we use (\ref{olp_xjdefead2}) to show (\ref{pnsnrlabcrt2}). From (\ref{olp_xjdefead2}) and the binomial distribution of $|u_{r,j}|$, for $r=2,\ldots,  r_n^{*}$, it holds that
\begin{align}
&  \mathbb{E}\scalebox{1.25}{\Bigg[} 1-\frac{\binom{X_{|u_{r,j}|,n}}{q}\binom{K_n}{q}}{\binom{P_n}{q}} \scalebox{1.25}{\Bigg]} \label{globesb}  \\
&  \leq  (1-p_n)^r + r p_n(1-p_n)^{r-1}{(1-s_n)} \nonumber \\
 &  \quad + [1-(1-p_n)^r -r p_n (1-p_n)^{r-1}] {(1-s_n)}^{1+\varepsilon_2}.   \label{bnmbcpinu1veead1ax2brst}
\end{align}
For $r=2,\ldots,  r_n^{*}$, based on (\ref{bnmbcpinu1veead1ax2brst}) and $t_n = p_n s_n$, we can further show
\begin{align}
 \text{(\ref{globesb})} & \leq  e^{-(1+\frac{\varepsilon_2}{2}){t_n}}.
 \label{eq:to_show2_crucial_bound_2stsa1}
 \end{align}
 Given (\ref{olp_xjdefead2}) and $0< \lambda_2 <1$, we obtain that $1-\frac{\binom{X_{|u_{r,j}|,n}}{q}\binom{K_n}{q}}{\binom{P_n}{q}}$ is upper bounded by $(1- s_n)^{ \lambda_2 |u_{r,j}|}$ for $|u_{r,j}| = 0, \ldots,  r_n^{*}$. Then it holds for $r=2, \ldots, r_n^{*}$ that
 \begin{align}
 \text{(\ref{globesb})} &\leq  \bE{(1- s_n)^{\lambda_2 |u_{r,j}|}}   = \big\{1-p_n[1-(1- s_n)^{\lambda_2}]\big\}^r. \nonumber
\end{align}
Then we   obtain  for $r=2, \ldots, r_n^{*}$ that
 \begin{align}
 \text{(\ref{globesb})} &\leq   e^{-\lambda_2 t_n r}, \label{pnsnrlab}
\end{align}
by deriving $ \big\{1-p_n[1-(1- s_n)^{\lambda_2}]\big\}^r    \leq  (1- {p_n} \cdot \lambda_2  s_n)^r  \leq e^{-\lambda_2 {p_n} s_n r}  = e^{-\lambda_2 t_n r}$,
where  we use $(1- s_n)^{\lambda_2} \leq 1 - \lambda_2 s_n$ due to $0 \leq s_n \leq 1$ and  $ 0<\lambda_2 < {\lambda}^q < \big(\frac{1}{2}\big)^q < 1$,    the fact that $1+x \leq e^x$ for any real $x$, and also $p_n s_n = t_n$.

 On the range $r=r_n^{*}+1, \ldots, n$, we establish
  \begin{align}
 \text{(\ref{globesb})}
&    \leq    \mathbb{E} \big[ {(1  -  s_n)^{\lambda_2 |u_{r,j}|}  \cdot  \1{|u_{r,j}|
\leq r_n^{*}}}\big]
\nonumber \\
 & \quad +  \mathbb{E} \big[ {e^{- \mu_2 K_n}   \cdot   \1{|u_{r,j}|
>   r_n^{*}} } \big] \nonumber \\
&   \leq     \mathbb{E} \big[  {(1 -  s_n)^{\lambda_2 |u_{r,j}|} } \big] +  e^{- \mu_2 K_n}
 \nonumber \\
 &   \leq  e^{-\lambda_2 t_n r} + e^{- \mu_2 K_n}  , \label{pnsnrlabcrt}
\end{align}
where the last step uses the result
 proved in
(\ref{pnsnrlab}).


The result (\ref{pnsnrlabcrt2}) is now proved given (\ref{eq:to_show2_crucial_bound_2stsa1}) (\ref{pnsnrlab}) and (\ref{pnsnrlabcrt}). Then as explained, we obtain  (\ref{ecnrbp1}), where the term $\bP{C_{n,r}} $ in (\ref{ecnrbp1}) is bounded below.

To bound $\bP{C_{n,r}} $, we let $\mathbb{G}_q(r)
$ be the subgraph of $\mathbb{G}_q
$ restricted to the vertex set $\{ v_1, \ldots , v_r \}$, and note   that $C_{n,r}$ means the event of $\mathbb{G}_q(r)
$ being connected. Let $\mathcal{T}_r$ denote the collection of all spanning trees on the vertex set $\{ v_1, \ldots , v_r \}$. We can   show for any $T \in \mathcal{T}_r$ that the probability of $T$ being a subgraph of $\mathbb{G}_q(r)
$ is ${t_n}^{r-1}$, where we recall $t_n$ as the edge probability in $\mathbb{G}_q
$. By Cayley's formula \cite{Martin}, there are $r^{r-2}$ spanning trees on $r$
vertices. This and the above result $\bP{T\subseteq\mathbb{G}_q(r)}={t_n}^{r-1}$ for any $T \in \mathcal{T}_r$, along with the union bound and $\bP{ C_{n,r}  } \leq 1$, together induce
\begin{align}
\bP{C_{n,r}}  \leq \min\{ r^{r-2}  {t_n}^{r-1},~1\}. \label{boundCnreq}
\end{align}

Applying (\ref{boundCnreq}) to (\ref{ecnrbp1}), we obtain 
 \begin{align}
  &  \bP{ A_{n,r,\ell} \bcap
\overline{E_n(\boldsymbol{X}_n)}}  \nonumber \\
 &   \leq \min\{ r^{r-2}  {t_n}^{r-1},~1\}   \times \min\{ (r t_n )^{\ell},~1\}  \times {g_{r,n}}^{n-r-\ell} . \label{a-2ndApplying}
\end{align}
The rest of the proof is using  (\ref{a-2ndApplying}) to prove (\ref{eq:OneLawToShow}). Due to space limitation, we provide the details in the full version \cite{fullversion}.
\hfill\fsquare

\section{Conclusion} \label{sec:Conclusion}

In this paper, we present a sharp zero--one law for secure $k$-connectivity
 in a  wireless sensor network under the
$q$-composite key predistribution scheme with unreliable links. Secure \mbox{$k$-connectivity}  ensures that any two sensors can find a path in between for secure communication even when at most $k-1$ sensors fail. The
network is modeled by composing a uniform \mbox{$q$-intersection}  graph with
an Erd\H{o}s--R\'enyi
graph, where the former characterizes the $q$-composite key predistribution scheme
and the latter captures link unreliability.

\linespread{1.1}

\end{document}